# Decomposition of low-angle grain boundaries

**Author information**

Wei Wan [1, 2, 3], Changxin Tang [2, *] and Eric R Homer [3, *]

[1] *Institute of Photovoltaics, Nanchang University, Nanchang, 330031, China*

[2] *Institute of Engineering Mechanics, Nanchang University, Nanchang, 330031, China*

[3] *Department of Mechanical Engineering, Brigham Young University, Provo, UT, 84604 USA*

\* *Corresponding authors, Email address:* tcx@ncu.edu.cn *(C.T.) and* eric.homer@byu.edu *(E.R.H.)*

**Abstract**

Grain boundaries (GBs) merge and grains disappear during microstructure evolution. However, the Peach-Koehler model predicts that particular stress states may reverse such a process by exerting differential Peach-Koehler forces on different dislocations. This work considers this reversal as GB decomposition and illustrates it in a low-angle asymmetric tilt GB and a low-angle mixed tilt-twist GB via atomistic simulation. In both cases, the dislocations separate into two GBs separated by a new grain. This work describes the requirements for decomposition and the importance of dislocation separability. Additionally, we examine the dislocation behaviors and stress signatures associated with this process, along with the impact of strain rate and temperature on those aspects.

**Introduction**

Grain boundary (GB) migration and its role in grain growth and microstructure evolution impact various behaviors of crystalline materials, such as strength, ductility, and corrosion resistance [1-6]. Decades of experimental and numerical progress have reported the migration mechanisms associated with various GB types, such as normal migration, shear coupling [7-12], sliding [13-15] and grain rotation [16-18]. In microstructure evolution, these mechanisms often result in the merging of GBs when grains disappear. However, no known evidence has confirmed the inverse process where a newly merged GB could "decompose" back to the original GBs that formed it.

The viability of such a process can be easily shown in a dislocation-structured, low-angle GB comprised of different dislocation types. The Peach-Koehler (P-K) force [19, 20], $F^k$, on each dislocation type $k$ (with the Burgers vector $b^k$ and line direction $l^k$) is calculated according to:

$$F^k = (\sigma \cdot b^k) \times l^k \tag{1}$$

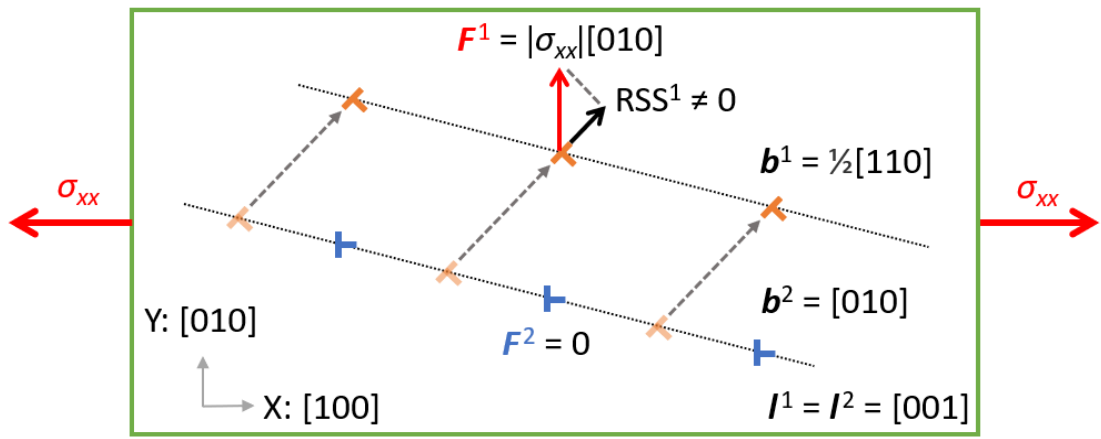

**Figure 1.** Model showing a [001] low asymmetric tilt GB separates into two asymmetric tilt GBs due to the different Peach-Koehler forces on dislocations.

In the case of the [001] low-angle asymmetric tilt GB shown in Figure 1, with two edge dislocation types, a uniaxial stress state, $\sigma = \sigma_{xx}$, results in $F^1 \neq 0$, and $F^2 = 0$. The resolved shear stress (RSS) can then be calculated for $F^1$ according to Schmid's law [20]. Since this stress state only results in the motion of dislocation type 1, the original boundary will decompose into two asymmetric tilt GBs with a new grain generated between them.

This ability to decompose one GB into two should be possible for any GBs that meet the following conditions: (i) the GB structure has dislocations with at least two different Burgers vectors, (ii) the stress state can induce sufficiently different Peach-Koehler forces for separation, and (iii) the dislocations involved must be able to separate from one another. These three criteria make this process unique as compared with previously reported GB dissociation [21, 22]. The asymmetric tilt GB in Figure 1 shows how all three criteria can easily be met. In contrast, mixed tilt-twist GBs could easily satisfy the first two criteria by combining the dislocation characteristics of both tilt and twist GB components [23-25]. But the third criterion of dislocation separability is more complicated because the tilt and twist GB components intersect. Nevertheless, since low-angle mixed GBs are often introduced in the production of quantum-dots and self-assembled nano-structures [26-29] and their migration behaviors are not well studied [30], decomposition could unveil new processing possibilities and applications.

In the present work, we employ molecular dynamics simulations to demonstrate GB decomposition in a low-angle asymmetric tilt GB (c.f. Figure 1) and a low-angle mixed tilt-twist GB. We examine the GB/dislocation behaviors under a particular stress state, as well as the temperature and strain rate dependence of the processes.

## Results & Discussions

### Asymmetric tilt GB

The actual response of the asymmetric tilt GB illustrated in Figure 1 shows the expected decomposition behavior when responding to the stress state described in the methods. As shown in Figures 2a1 and 2a2, the [010] dislocation does not migrate while the ½[110] dislocation does migrate. The ½[110] dislocation array gradually glides away from the [010] dislocation, resulting in two asymmetric tilt GBs illustrated in Figure 2a3. This confirms the expected decomposition based on the predicted behaviors from the Peach-Koehler model. It is noteworthy that this decomposition is reversible because the GB-dislocation structures interact only through forces and the stress-strain curve in Figure 2b transits smoothly. The reversibility is an important signature that distinguishes this kind of decomposition from those in the mixed GBs.

### Low-angle mixed grain boundary

We also consider the response of a more complex low-angle, mixed GB presented in Figure 3a1. Its tilt component is a (001)/[100] low-angle symmetric tilt GB comprised of a [001] edge dislocation array. The twist component is a (001) low-angle, twist GB comprised of a ½[110] screw dislocation network. The mixed GB itself has a dislocation structure where the edge dislocation arrays and the screw dislocation network intersect at select nodes. At the intersection node, the two dislocation types react to form two small loops, one slightly bigger than the other, where the edge dislocation previously existed; this is shown in detail in *Supplementary* Figure S1. This small but important change in structure plays an important role that is discussed later.

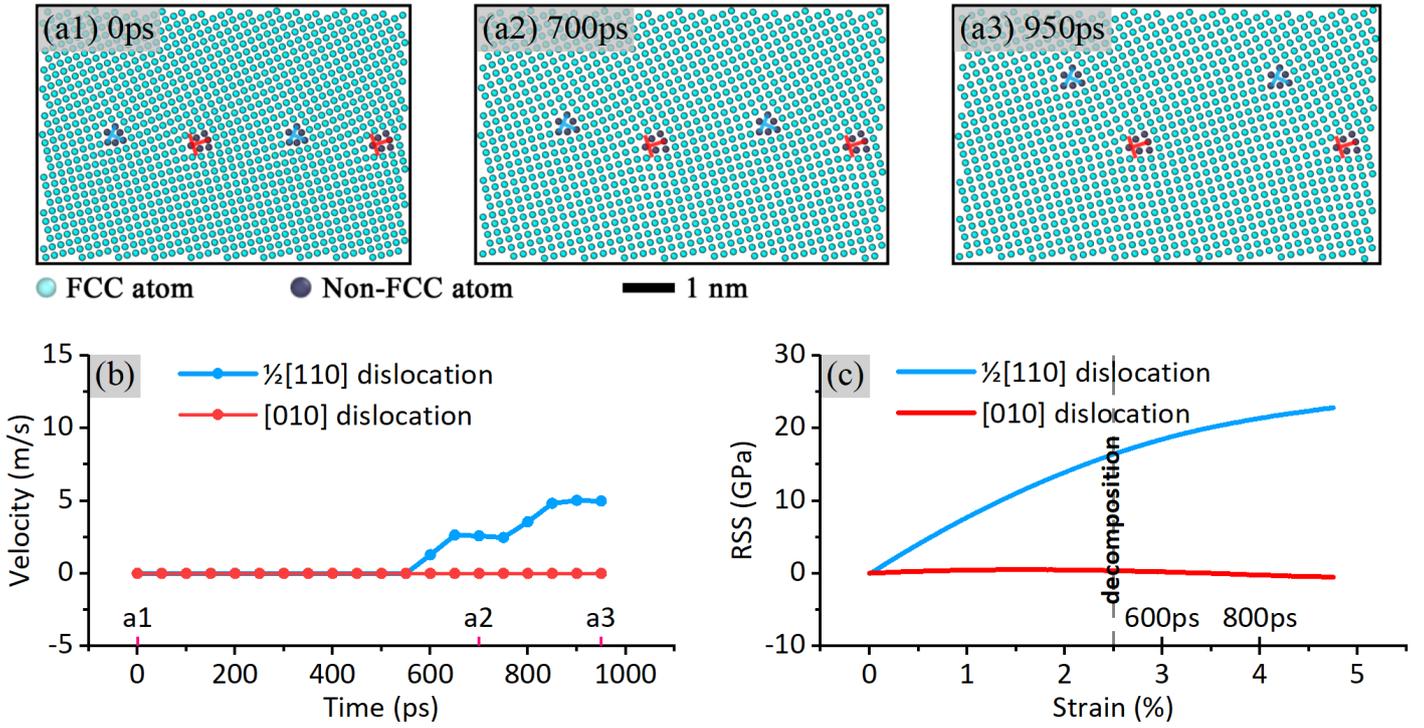

**Figure 2.** Decomposition of an asymmetric tilt GB deformed at $5\times10^7$ s$^{-1}$ and 100 K. (a1) Initial GB structure at 0 ps; (a2) GB structure at 700 ps; (a3) GB structure at 950 ps; Conjugate gradient energy minimization is used to remove the thermal noises in (a1), (a2) and (a3); (b) Velocities of the two dislocation types in the low-angle asymmetric tilt GB; (c) RSS-strain/time curve of the two dislocation types. It is also noted that after 950 ps, the applied stress results in fracture of the simulation cell and additional motion is not possible.

The stress state, as described in the methods, is designed to induce motion of the screw dislocations while keeping the edge dislocations stationary. The decomposition process is presented in Figure 3. The violin plot in Figure 3a characterizes the distributions of GB atoms thereby tracking the position variations of the two GB components (i.e. separation of the GB components leads to multiple peaks in the violin plot). At 600 ps, the screw dislocation network has bowed out while the edge dislocation array stays static (c.f. Figure 3a2). This configuration remains largely unchanged for several hundred picoseconds because the screw dislocation is essentially pinned by the stationary edge dislocations. During this time, the stress-strain curve in Figure 3b shows that the RSS increases linearly for the screw dislocation but remains near zero for the edge dislocation. The decomposition finally occurs at 1600 ps, as shown in Figure 3a3, where the screw dislocation network separates from the stationary edge dislocation array and restores to a nearly flat low-angle twist GB. Hence, the original low-angle mixed GB decomposes to independent tilt and twist GBs with a new grain formed between the two. Figure 3c illustrates how the RSS impacts the GB position for each dislocation type/GB component.

In order to better understand the response of the mixed GB, which contains co-existing tilt and twist GB components, we examine and contrast the independent response of the two GB components separately, referred to as isolated tilt and twist GBs. Under the same conditions, the isolated tilt GB is stationary and the isolated twist GB is mobile, as shown in Figures 3d, 3d1 and 3d2. Comparison of the RSS-strain curves in Figures 3e and 3b suggests that there is a stress barrier that must be overcome for the screw dislocations to separate from the edge dislocation. This barrier appears to be ~0.12 GPa above the 0.3 GPa stress (Figure 3f) required for the isolated twist GB to migrate under these conditions. One can view this as the barrier that had to be overcome to allow the GB to decompose (i.e., the barrier to allow the screw dislocation network to separate from the edge dislocations in the mixed GB).

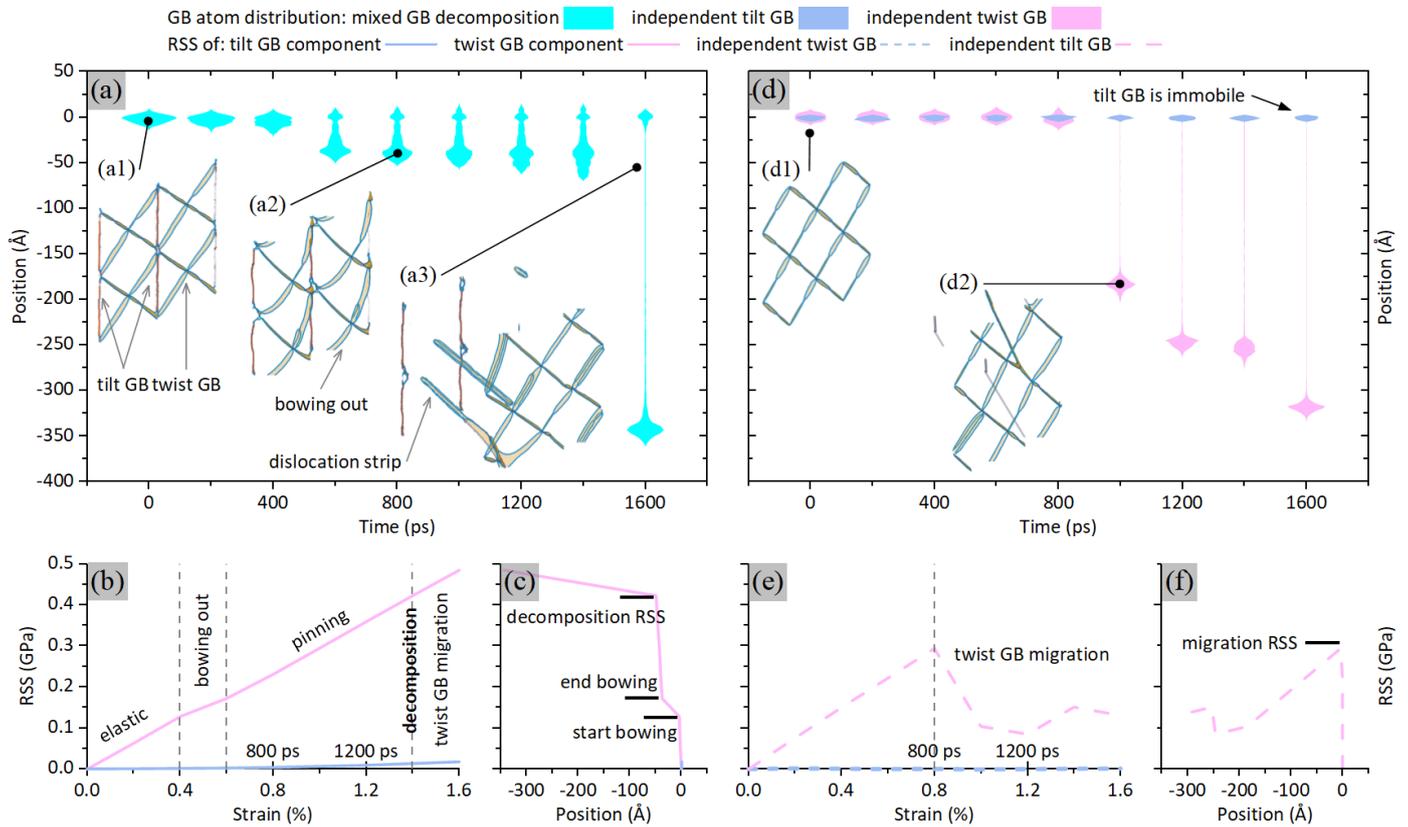

**Figure 3.** Decomposition process of the mixed GB as contrasted with the migrations of its two independent GB components, at 100 K and $10^7$ strain rate: (a) Violin plot showing the GB atom distribution (non-FCC atoms identified by adaptive common neighbor analysis [31]) as a function of both time and position along the boundary plane normal; (a1) Initial mixed GB structure at 0 ps; (a2) Mixed GB structures with the bowing screw dislocations at 800 ps; (a3) Decomposed structures at 1600 ps; (b) RSS-strain/time curve; (c) RSS versus position for the twist GB component. (d) Violin plot showing the distributions of tilt and twist GB atoms as a function of both time and position; (d1) Initial twist GB structure at 0 ps; (d2) Twist GB structures at 800 ps; (e) RSS-strain/time curve and (f) RSS versus GB positions, for the migrations of the isolated tilt and twist GBs; the isolated tilt GB remains stationary.

One may also wonder if the decomposition is a reversible process, since merging a pair of tilt and twist GB is theoretically easy to access. Unfortunately, for any low-angle mixed GBs, the decomposition must break the dislocation network nodes [32, 33] and leave dislocation "strips" that would fundamentally change the GB structures. These strips and their population in Figure 3a3 suggest that cyclic decomposition and merging would be very difficult.

**Strain rate and temperature effects**

As noted in the introduction, decomposition of a given GB appears to require satisfaction of three conditions: (i) dislocations with at least two different Burgers vectors, (ii) a stress state that causes differential

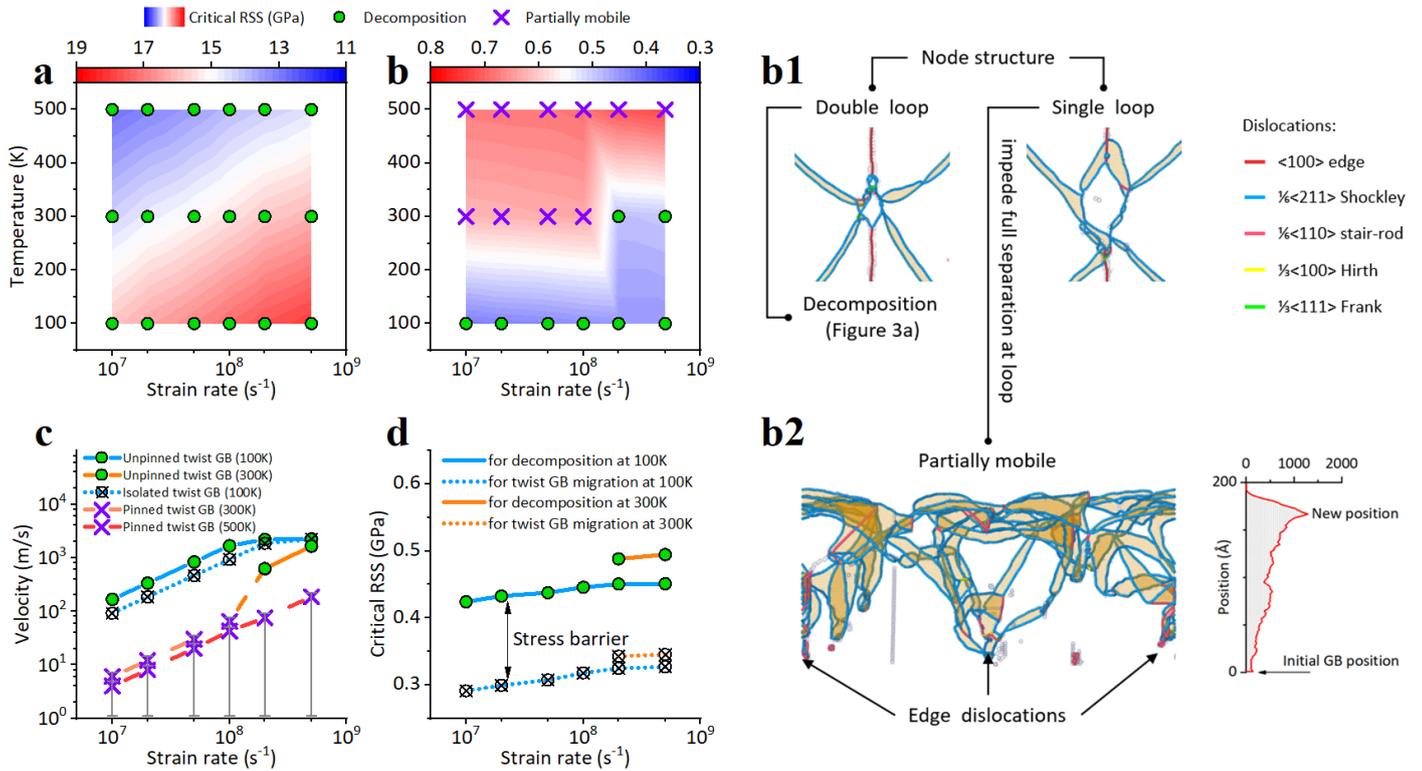

**Figure 4.** (a) Critical RSS for the decomposition of the low-angle asymmetric tilt GB. (b) Critical RSS for the decomposition and partially mobility of the low-angle mixed GB. (b1) Node structures taken at the same strain that precede decomposition or partially mobility. (b2) A partially mobile mixed GB structure with many sessile dislocations as the pinning point. (c) GB velocity as the function of strain rate and temperature, for the cases of decomposition and partially mobility. Isolated twist GB velocities are provided for contrast. (d) Comparison of critical RSS for the mixed GB decomposition and isolated twist GB migration reveals the additional stress barrier to separate a twist GB at both 100K and 300K for all strain rates.

Peach-Koehler forces on the dislocations, and (iii) separability of the dislocations. As discussed above, the asymmetric tilt GB easily achieves the three conditions while the mixed tilt-twist GB only achieves these conditions after stress has increased and time has elapsed. As dislocation migration and reactions are known to be temperature and strain-rate dependent [34], we examine these two GBs at a range of strain rates and temperatures, as described in the methods. Figures 4a and 4b contain details of the low-angle asymmetric tilt GB and mixed GB, respectively. Each plot contains markers at the examined strain rates and temperatures; the markers denote whether the GB decomposes or not, marked with a green circle or purple ✕, respectively. Additionally, the plot contains contours of critical RSS. For the asymmetric tilt GB, the critical RSS is measured when the mobile dislocations start to migrate. For the mixed GB, the highest RSS achieved during each simulation is reported, which is considered as the critical RSS to activate decomposition if it occurs or to cause partial mobility if decomposition doesn't occur.

Unsurprisingly, decomposition always occurs for the low-angle asymmetric tilt GB since the two dislocation types are not physically interacting or pinning one another. In this GB, the critical RSS increases with the strain rate and decreases with increased temperature. This is characteristic of a thermally activated process. Higher temperatures lower the required driving force for a given process. Higher strain rates yield less time for a random fluctuation to occur, leading to higher driving forces. Thus, it appears that even though these two dislocations do not intersect, there is a large driving force required to separate these dislocations and induce decomposition.

For the low-angle mixed GB, decomposition is only observed at low temperatures and high strain rates. When decomposition doesn't occur, the screw dislocation network bows out and can even migrate to some degree. However, it is still pinned by the stationary edge dislocations; during the elapsed time of the simulations, the screw dislocation never fully separate (c.f. Figure 4b2). Based on the fact that a part of the mixed GB is pinned while the rest moves in a non-uniform, dislocation-type-dependent velocity, we name this behavior partially mobile and discuss it briefly in the Supplementary Materials. Figure 4b also illustrates just how much higher the RSS is to cause the mixed GB to be partially mobile as compared with the RSS to induce decomposition.

Inspection of the dislocation structures before full or partial separation of the screw dislocation shows two different node structures, which are presented in Figure 4b1. In both cases, the edge dislocation is not present at the location where the two dislocations should intersect because the two dislocation types have reacted to form new node structures. In the case of GBs that decompose, the reaction forms a double loop where the edge dislocation is eliminated, whereas in GBs without decomposition, the double loop becomes a single loop. Supplementary Figure S2 shows these node structures for all examined strain rates and temperatures. Dislocation analysis tool (DXA) [35] shows that both structures in this state can contain sessile dislocations, such as the stair-rod (signature of a Lomer-Cotrell Lock), Frank, and Hirth dislocations [20]. However, as shown in Supplementary Figures S3 and S4, the node reactions with the double loop later allow the screw dislocation to break away, but those with single loops are unable to break away. It is hypothesized that the line tension from the bowing screw dislocations (c.f. Figure 3a2) forces some of these reactions to reverse to allow the screw dislocations to break away from the intersection node. But, in the case of the single loop node structure, the reactions are difficult to reverse to allow the screw dislocations to break away. As noted earlier, an example of the resulting structure where the screw dislocation is unable to break away is illustrated in Figure 4b2; these networks become complex and contain numerous segments of sessile dislocations.

The hypothesized reasoning behind why decomposition is observed at low temperatures and high strain rates and not in the alternative cases is as follows. Dislocation reactions are facilitated by thermal energy [36]. At low temperatures, the thermal energy appears to be insufficient for more complex reactions to occur at these intersection nodes. Therefore, at the lowest temperature of 100 K, the double loop node structure that forms never evolves further and sufficient applied stress leads to breakaway of the screw dislocations (c.f. Supplementary Figures S2 and S3). At 300 K, decomposition is observed at low strain rates but not at high strain rates. It is hypothesized that the slower strain rate provides more time in the presence of additional thermal energy for the double loop node structure to evolve into the single loop node structure (c.f. Supplementary Figures S2 and S4). Under higher strain rates, less time elapses before the hypothesized line tension of the bowed screw dislocations leads to break away of the screw dislocation; the node structure does not evolve from the double loop structure due to insufficient time. At 500 K, the available thermal energy is sufficient that only the single loop node structure is observed and the screw dislocations never fully break away from the edge dislocation.

A comparison of dislocation velocities normal to the original GB plane helps to illustrate the difference between the mixed GBs that have undergone decomposition and those that are only partially mobile. The dislocations in the case of decomposition are the screw dislocation network of the twist GB component, referred to as unpinned twist GB. In the case of the partially mobile network, it contains portions of both screw and reacted edge dislocations (c.f. Figure 4b1), referred to as pinned twist GB. Since the dislocations in both these cases do not migrate at constant velocity, the method used to extract the velocities from the simulations is described in the *Supplementary Materials*. For reference, we also examine the migration and velocity of the isolated twist GB, referred to as isolated twist GB. Figure 4c plots the velocities of all three twist GB types as a function of strain rate. These data are instructive to illustrate the rate and temperature dependence of the twist GB (screw dislocation network) under different structural states. We note four features in the plot. (1) The isolated and unpinned twist GBs move faster than the pinned twist GB. This is unsurprising since the pinned twist GB is only partially mobile due to pinning in some locations and needing to drag along some of the reacted tilt components. (2) The unpinned twist GB moves faster than the isolated twist GB, likely due to the higher stress built up prior to its separation (c.f. Figures 3c and 3f). (3) The isolated and unpinned twist GBs do not continue the linear relationship between velocity and strain rate at the highest high strain rates. This phenomenon has been shown previously when dislocation velocities approach forbidden wave speeds [37]. (4) At 300 K, one can see the changes from slow migration at lower strain rates for the pinned twist GB to fast migration at higher strain rates for the unpinned twist GB. Despite significant differences in mechanisms, this is similar to the phenomenon of solute drag where a GB will migrate slowly at low driving forces and faster when it breaks away at higher driving forces [38-41].

The concept of a stress barrier presented earlier appears to be consistent for the low-angle mixed GB across the conditions where decomposition is observed. The critical RSS for the unpinned and isolated twist GBs are illustrated in Figure 4d as a function of strain rate. It can be seen that the stress barrier, calculated as the difference between the two, remains nearly constant with the strain rate and is consistent with the velocity differences shown in Figure 4c. The stress barrier for decomposition of the low-angle mixed GB, expressed as an average of the difference between the two values and with uncertainty of the standard deviation of the difference, has an apparent value of $134 \pm 10$ MPa.

## Conclusions

The Peach-Koehler model predicts that single GBs are capable of decomposing into two different GBs under particular stress states if their structure is comprised of at least two different and separable dislocations. We have demonstrated this idea via atomistic simulation in two low-angle GBs, one asymmetric tilt and one mixed tilt-twist.

The conclusions are summarized as follows: (1) The stress state is controlled to induce migration in only one dislocation. For the asymmetric tilt GB, one edge dislocation migrates while the other does not. For the mixed GB, the screw dislocation network (twist component) migrates while the edge dislocation array (tilt component) remains static. However, before it could migrate independently, the screw dislocation network bowed out under the stress due to the constraint of the stationary edge dislocations to which it was connected. After reaching a critical stress, it separated from the edge dislocation array and quickly restored to a nearly planar structure. In both cases, the original GB becomes two independent GBs defined by the two dislocations that separated under the applied stress, with a newly formed grain between them.

(2) The decomposition of the GBs is subject to the separability of the dislocations. For the low-angle asymmetric tilt GB, this is readily achieved for the two unconnected edge dislocations. For the low-angle mixed GB, the separability is affected by the dislocation pinning effects at the intersection nodes. These dislocation pinning effects are strain-rate and temperature dependent. Decomposition can be achieved if the intersection node structure allowed the screw dislocations to separate from the edge dislocations. At higher temperatures and lower simulation times (slower strain rates) the node structure evolves to make decomposition difficult. Although the screw dislocations are still driven by the applied stress to migrate, some of its segments are pinned by the stationary edge dislocations and the dislocation network is only partially mobile. This leads to slower dislocation velocity and increased critical RSS necessary to induce migration.

As noted in the introduction, low-angle GBs impact a variety of relevant fields like self-assembled nano-templates and quantum dots. Decomposition could provide a mechanical alternative to control the dislocation pattern in thin films, instead of precise disorientation angle control [27-29]. Furthermore, if designed in an appropriate way, GB decomposition could be used to sweep dislocation arrays through grains to gather defects or impurities [42] in critical applications like nuclear materials, etc.

Finally, we have demonstrated the viability of decomposing low-angle GBs comprised of dislocation structures here, but we also note that high-angle GBs can decompose. Additional work by some of the authors [43] shows the decomposition of high-angle GBs, computationally and experimentally, through disconnection mechanisms. The discovery of a new grain that appears to have emerged from an existing GB during annealing in the dataset by Bhattacharya, et al. [44], as described in [43], suggests that this phenomenon may already be active but sufficiently rare that it has not previously been reported.

**Methodology**

Based on the crystal structure, the selection of materials with the same lattice type should not affect the dislocation structures of the given low-angle GB character. We consider the low-angle GBs in FCC nickel as an example to investigate the predicted decomposition process, because dislocations in FCC metals are relatively easy to glide, compared with other lattice types. We use the EAM potential contributed by Foiles and Hoyt [45] because it is widely used in the simulation of nickel GB migration.

A low-angle asymmetric tilt GB with good visualization effects is used as the toy model, where the disorientation/tilt axis is [010], the disorientation angle is 8.8° and the asymmetric tilt angle is 18.43°. The low-angle mixed GB is constructed following the methodology stated in our previous work, where the boundary plane is close to (001) and the disorientation axis is [100] and the disorientation angle is 1.35°. Additional details about simulation (size, etc.) are provided in the *Supplementary Materials*. The structures of the two low-angle GBs are generated under zero temperature and pressure by the LAMMPS software [46] and strictly following the published sampling method [47, 48]

Once the GB structures are generated, the DXA tool [35] integrated in OVITO software [49] is used to determine the Burgers vector and line direction of the dislocations in the low-angle GBs. Based on the identified structural information, the Peach-Koehler model is used to determine the desired stress states to induce decomposition for each GB, which are described in *Supplementary Materials*. We impose constant strain boundary conditions that lead to the desired stress states and keep the strain rate constant during the strain loading. Atomistic simulations are performed at zero pressure and temperatures of 100 K, 300 K and 500 K, combined with commonly used molecular dynamic simulation strain rates $10^7$, $2\times10^7$, $5\times10^7$, $10^8$, $2\times10^8$ and $5\times10^8$ s$^{-1}$. The simulations are run to achieve the same total strain for each strain rate.

The RSS of the dislocation $k$ is calculated by considering the full six global stress components $\sigma_{ij}$ of the simulation cell following:

$$\text{RSS}^k = \boldsymbol{b}_j^k \boldsymbol{\sigma}_{ij} \boldsymbol{n}_i^k, \tag{2}$$

where $\boldsymbol{n}$ is the normal vector of the slip plane of dislocation $k$.

## Data availability

Numerical data are available upon reasonable request.

## Acknowledgements

W. Wan acknowledges the insightful discussion of the grain boundary dissociation with Prof. Jingbo Yang from Ningbo Institute of Technology, Zhejiang University. W. Wan and E.R. Homer acknowledge insightful discussions with Ethan Cluff and Prof. Oliver K. Johnson from Brigham Young University.

## Competing interests

The authors declare no competing interests.

## Fundings

E.R. Homer was supported by the U.S. Department of Energy (DOE), Office of Science, Basic Energy Sciences (BES) under award DE-SC0016441. W. Wan acknowledges the financial support from (1) the Department of Mechanical Engineering, Brigham Young University; (2) National Natural Science Foundation of C.X. Tang. C.X. Tang was supported by the National Natural Science Foundation of China (grant number: 12464044).